\documentclass[reprint,a4paper,prstab,superscriptaddress]{revtex4-2}

\usepackage[colorlinks=true,linkcolor=blue,citecolor=blue,urlcolor=blue]{hyperref}
\usepackage[pdftex]{graphicx}
\usepackage{siunitx}
\usepackage{amsmath}
\renewcommand{\t}{\text}

\renewcommand{\b}{\boldsymbol}
\newcommand{\bu}{\mathbf}
\newcommand*\diff{\mathop{}\!\mathrm{d}}
\newcommand{\iu}{\ensuremath{\mathrm{i}}}
\newcommand{\eu}{\ensuremath{\mathrm{e}}}

\begin{document}

\title{Spectral narrowing of a phonon resonance in time-domain sum-frequency spectroscopy}

\author{R. Kiessling}
\author{M. Wolf}
\author{A. Paarmann}
\email{alexander.paarmann@fhi-berlin.mpg.de}
\affiliation{Fritz Haber Institute of the Max Planck Society, Faradayweg 4-6, 14195 Berlin, Germany}

\begin{abstract}

Sum-frequency generation (SFG) spectroscopy provides a versatile method for the investigation of non-centrosymmetric media and interfaces. Here, using tunable picosecond infrared (IR) pulses from a free-electron laser, the nonlinear optical response of 4H-SiC, a common polytype of silicon carbide, has been probed in the frequency- and time-domain by infrared-visible vibrational SFG spectroscopy. In the SFG spectra we observe a sharp resonance  near the longitudinal optical phonon frequency, arising from linear optical effects due the epsilon-near-zero regime of the IR permittivity. In the time domain, the build-up of the SFG intensity is linked to the free-induction decay of the induced coherent IR polarization. When approaching the frequency of the phonon resonance, a slower polarization dephasing is observed as compared to off-resonant IR excitation. Thus, by introducing a temporal delay between the IR and the visible up-conversion pulse we are able to demonstrate spectral narrowing of the phonon SFG resonance, as corroborated by model calculations.

\end{abstract}

\maketitle

\section{Introduction}

Lattice vibrations, as an elementary excitation of condensed matter, play a fundamental role for the properties of solids. Thus, investigations of vibrational modes yield insights into the dynamical behavior of atomic or molecular systems at the microscopic scale~\cite{Fri07,Eic10,Foer11}. 
In this regard, infrared-visible (IR-VIS) sum-frequency generation (SFG) spectroscopy has proven to be a valuable all-optical method~\cite{She14}. 
Besides information about the energy and spectral lineshape of the vibrational excitations, the nonlinear optical character of the technique gives also access to symmetry properties of the investigated system~\cite{She89,She16}. In particular, interface- and surface-specific spectra of vibrational modes, e.g. of surface-adsorbed species, can be studied~\cite{Vid05,Nih09}. In addition, time-resolved SFG spectroscopies directly elucidate the energy and phase relaxation dynamics of the vibrational states~\cite{Bon00,Arn10}.\\

Whereas IR pump-SFG probe spectroscopy is utilized to measure the lifetime $T_1$ of a vibrational population, the dephasing time $T_2$ of the induced polarization can be directly obtained from time-domain SFG spectroscopy~\cite{Ueb97}.  There, the IR pulse excites a coherent polarization whose decay is subsequently probed by the time-delayed VIS pulse. The measured phase relaxation, also called free-induction decay (FID), is linked to the SFG spectrum via Fourier transform~\cite{Hes02}.
Thus, deviations in the FID affect the spectral response and vice versa. Previous IR-VIS SFG studies have investigated the effect of linear and nonlinear optical properties on the static spectra of solid materials~\cite{Bar94,She08} as well as both the frequency- and time-domain response of adsorbate vibrations~\cite{Bon00} and interfacial systems~\cite{Rok03,Bor05}.\\ 

Here, the SFG response in the spectral range of the optical phonons in silicon carbide, a robust semiconducting crystal with a broad range of applications~\cite{Fis17}, has been investigated. 
Besides the typical frequency-domain SFG spectroscopy, also time-domain SFG measurements are performed. The experimental data allow to analyze the relation between the free-induction decay and changes in the transient SFG spectra, supported by simulations of the time-domain SFG response. Finally, consequences for the spectral resolution in time-domain SFG experiments are discussed.

\section{Experimental System}

The spectral-domain and time-domain nonlinear optical response of silicon carbide is investigated by IR-VIS SFG spectroscopy. This technique employs a tunable, narrowband IR pulse with frequency $\omega_{1}$ for excitation of the vibrational resonance, while the delayed up-conversion pulse in the VIS spectral range (frequency $\omega_{2}$) leads to the emission of a photon at the sum frequency $\omega_3 = \omega_1 + \omega_2$. By scanning the IR frequency $\omega_1$ at constant up-conversion field $\omega_2$, the detected SFG intensity $I(\omega_3)$ reveals the nonlinear vibrational spectrum of the investigated system. In contrast, tuning of the time delay $\tau$ between both pulses at fixed frequencies while recording $I(\tau)$ gives direct access to the time-domain dynamics of the induced IR polarization~\cite{Hes02}.\\

The coherent IR radiation with a continuously tunable center wavelength between $3$ and $\SI{50}{\micro m}$ has been obtained from a free-electron laser (FEL) oscillator~\cite{Schoe15}. There, the repetition rate of the accelerated electron bunches is imprinted on the temporal structure of the IR emission, delivering ps-short FEL micro-pulses at a rate of $\SI{55.5}{MHz}$ within a $\sim\SI{10}{\micro s}$-long macro-pulse. The duration, and thus bandwidth, of the micro-pulses can be adjusted via shortening of the FEL cavity length $L$, described as the cavity detuning $\Delta L$~\cite{Kie18}.\\

\begin{samepage}
The VIS up-conversion radiation is derived from a table-top NIR fiber oscillator, providing bandwidth-limited $\SI{100}{fs}$-short pulses at $\SI{1055}{nm}$ wavelength and $\SI{55.5}{MHz}$ repetition rate. Half of the NIR power is frequency-doubled in a $\SI{1}{mm}$ thick, type I phase-matched BBO crystal, whereas the remaining light is used for a balanced optical cross-correlator (see below). For the measurement of the static SFG spectra, the VIS pulses are stretched in a strongly dispersive SF11 glass to about $\SI{1}{ps}$ duration (FWHM), increasing the temporal overlap with the IR FEL pulses. In the time-domain experiments, the original $~\SI{120}{fs}$ VIS pulses pass a delay stage before being focused onto the sample. The temporal overlap of IR and VIS pulses is accomplished by an RF synchronization system between the FEL master oscillator and the table-top laser, see Ref.~\cite{Kie18} for details.

\end{samepage}

Static SFG spectra are acquired within a reflection geometry. There, the angle of incidence (AOI) of the IR radiation is $\SI{55}{\degree}$, whereas the VIS light is incident under $\SI{30}{\degree}$. Due to momentum conservation, the generated sum-frequency radiation is emitted close to the propagation direction of the reflected VIS beam. Spectral and polarization filtering is used before SFG intensity detection by a photomultiplier tube. The linear polarization direction of IR and VIS radiation can be adjusted via half-wave plates. 
In the time-domain spectroscopy measurements, the sample is mounted with its surface normal parallel to the VIS wavevector and the transmitted SFG radiation is detected. In parallel, a balanced optical cross-correlator (BOC)~\cite{Sch03} monitors the actual FEL -- table-top laser pulse timing. For that, a part of the NIR radiation is mixed with $\SI{15}{\percent}$ of the FEL power in a GaSe crystal for phase-matched sum-frequency generation. The determination of the timing correction value $\delta \tau$ by the BOC method is detailed in Ref.~\cite{Kie18}. The measured time-domain SFG traces $I(\tau)$ shown below have been post-corrected by the recorded BOC timing $\delta \tau$ as described in the Appendix.\\

The investigated sample is a $\SI{350}{\micro m}$ thin, $c$-cut semi-insulating silicon carbide crystal of polytype 4H. Due to the point group symmetry $6mm$~\cite{Hof94}, the hexagonal crystal exhibits birefringent linear optical properties. Further, the lack of an inversion center allows for a nonzero electric dipole contribution to a strong second-order nonlinear optical susceptibility $\chi^{(2)}$~\cite{Nie99,Sat09}. The wide band gap of $E_\t{g}=\SI{3.2}{eV}$~\cite{Fis17} inhibits electronic excitations by visible photons, while the strong Si-C bonds within the large crystal unit cell give rise to multiple vibrational modes in the mid-infrared spectral region~\cite{Hof94}, causing the so-called Reststrahlen band~\cite{Paa15}. All spectroscopic measurements have been performed at ambient conditions.

\section{Static Sum-Frequency Spectrum}

\subsection{Theoretical Model}

For polar dielectric materials, the sum-frequency response is dominated by the electric-dipole contribution from the bulk phase. Thus, the spectral behavior of the generated SFG intensity is strongly influenced by the dispersion of both the linear and second-order nonlinear optical susceptibilities. Within the established theoretical model, the SFG signal is given by~\cite{She08,She16}
\begin{multline}
	\label{eq:SFGbulk}
	I(\omega_3) \propto \left| \bu{F}(\omega_3) \hat{\b{e}}(\omega_3) \cdot \b{\chi}^{(2)}_\t{B} :  \bu{F}(\omega_1) \hat{\b{e}}(\omega_1) \; \bu{F}(\omega_2) \hat{\b{e}}(\omega_2) \right|^2 \\ \cdot 1/ \Delta k^2,
\end{multline}
with the Fresnel factor $\bu{F}(\omega)$, the bulk susceptibility $\b{\chi}^{(2)}_\t{B}$ and the wavevector-mismatch $\Delta k$. In the case of optical anisotropic media, the dielectric function $\epsilon(\omega)$ as well as the second-order susceptibility $\b{\chi}^{(2)}_\t{B}$ obtain tensorial character, where symmetry properties of the crystal structure limit the number of independent non-vanishing components. The polarizations of the incident and emitted electric fields $\b{E}(\omega_i)$ are accounted for in the nonlinear response function, Eq.~(\ref{eq:SFGbulk}), by the unit vectors $\hat{\b{e}}_\t{p}(\omega_i) = (\cos \theta_i,0,\sin \theta_i)$ or $\hat{\b{e}}_\t{s}(\omega_i) = (0,1,0)$ for $p$- or $s$-polarized waves, respectively. The angle of incidence or emission of field $\b{E}(\omega_i)$ relative to the surface normal is denoted by $\theta_i$.\\

The Fresnel factors describe the behavior of an optical field transversing the interface of two different media. For the case of an isotropic medium (labeled I) next to an uniaxial crystal (II) whose surface normal is parallel to the optic axis along $z$-direction ($c\parallel z$), the Fresnel coefficients are~\cite{Mor18,Paa15}
\begin{equation}
	\label{eq:Fresneltensor}
	\begin{aligned}
		F_{xx} &= \frac{2 \varepsilon^\t{I} g_\t{e}^\t{II}}{\varepsilon_\perp^\t{II} g^\t{I} + \varepsilon^\t{I} g_\t{e}^\t{II}},\\
		F_{yy} &= \frac{2 g^\t{I}}{g^\t{I} + g_\t{o}^\t{II}},\\
		F_{zz} &= \frac{\varepsilon_\perp^\t{II}}{\varepsilon_\parallel^\t{II}}  \frac{2 \varepsilon^\t{I} g^\t{I}}{\varepsilon_\perp^\t{II} g^\t{I} + \varepsilon^\t{I} g_\t{e}^\t{II}}.
	\end{aligned}
\end{equation}
The dispersion functions of the anisotropic medium are part of the dielectric tensor with $\varepsilon_{xx}(\omega), \varepsilon_{yy}(\omega) = \varepsilon_\perp(\omega)$ and $\varepsilon_{zz}(\omega) = \varepsilon_\parallel(\omega)$. The relevant $z$-components of the wavevector $g \equiv k_z$ read
\begin{equation}
	\label{eq:kz}
	\begin{aligned}
		g^\t{I}(\omega,\theta) &= \frac{\omega}{c_0} \sqrt{\varepsilon^\t{I}(\omega)} \cos \theta,\\
		g_\t{o}^\t{II}(\omega,\theta) &= \frac{\omega}{c_0} \sqrt{\varepsilon_\perp^\t{II}(\omega) - \sin^2 \theta},\\
		g_\t{e}^\t{II}(\omega,\theta) &= \frac{\omega}{c_0} \sqrt{\frac{\varepsilon_\perp^\t{II}(\omega)}{\varepsilon_\parallel^\t{II}(\omega)}} \sqrt{\varepsilon_\parallel^\t{II}(\omega) - \sin^2 \theta},
	\end{aligned}
\end{equation}
with the angle of incidence $\theta$ relative to the surface normal. The birefringence of the uniaxial crystal is taken into account by distinguishing the propagation of ordinary and extraordinary beam, denoted as $o$ and $e$, respectively.\\

The phonon resonances in the linear optical function of the crystal are modeled as Lorentz-type oscillators. Within the infrared spectral range, where the Reststrahlen effect is observed, the following function is used~\cite{Mut99},
\begin{equation}
	\label{eq:eps_sic}
	\varepsilon_{p}(\omega_1) = \varepsilon_{p,\infty} \left(1 + \sum_r \frac{\omega^2_{\t{LO}\,r,p} - \omega^2_{\t{TO}\,r,p}}{\omega^2_{\t{TO}\,r,p} - \omega_1^2 - \iu\gamma_{\t{TO}\,r,p}\,\omega_1} \right)
\end{equation}
with the high-frequency dielectric constant $\varepsilon_\infty$, the frequencies~$\omega_{\t{TO}\,r,p}$ and dampings~$\gamma_{\t{TO}\,r,p}$ of the IR-active TO phonon resonances $r$. The numerator $\omega^2_{\t{LO}\,r,p} - \omega^2_{\t{TO}\,r,p}$ represents the oscillator strength. Each quantity is given with respect to the basal plane ($p =\, \perp$) and parallel to the $c$ axis~($p =\, \parallel$) of the anisotropic crystal, respectively.\\

In the same way, the dispersion of the second-order susceptibility tensor components $\chi^{(2)}_{ijk}(\omega_3,\omega_2,\omega_1)$ is modeled as sum of a constant contribution and an IR-frequency dependent Lorentz oscillator for each SFG-active phonon resonance~\cite{Fau66}, i.~e.
\begin{equation}
	\label{eq:FH}
	\chi^{(2)}_{ijk}(\omega_3,\omega_2,\omega_1) = \chi^{(2)}_{ijk\,\infty} \left(1 + \sum_{r} \frac{A_{r,ijk}}{\omega_{\t{TO},r}^2 - \omega_1^2 - \iu \gamma_{\t{TO},r} \omega_1} \right),
\end{equation}
which peaks at each IR-active TO phonon resonance $r$, with $A_{r,ijk}$ being the associated Faust-Henry coefficient~\cite{Fau66}. Due to the optical frequencies $\omega_1, \omega_2, \omega_3 < E_\t{g}/\hbar$ being well below the material's band gap, no electronic resonances are expected, i.e., $\chi^{(2)}_{ijk}$ only depends on the IR frequency $\omega_1$.\\

The number of independent components of the $\b{\chi}^{(2)}_\t{B}$ tensor is governed by the point group symmetry of the crystal~\cite{Boy08}. Usage of the second-order tensor in accordance with the experimental arrangement requires a transformation from the crystallographic framework $(i,j,k) \in (a,b,c)$ with the principal axes $a,b$ and $c$ to the laboratory coordinates $(l,m,n) \in (x,y,z)$ as given by
\begin{equation}
	\label{eq:transform}
	\chi^{(2)}_{lmn} = \sum_{ijk} \chi^{(2)}_{ijk} (\b{e}_l \cdot \b{e}_i) (\b{e}_m \cdot \b{e}_j) (\b{e}_n \cdot \b{e}_k).
\end{equation}

The wavevector mismatch $\Delta k = |\b{k}(\omega_3) - \b{k}(\omega_2) - \b{k}(\omega_1)|$ inside the investigated material extends along the $z$-direction ($\Delta k = \Delta k_z = \Delta g$) since the momentum parallel to the interface is conserved. In general, the wavevector $k(\omega_i)$, Eq.~(\ref{eq:kz}), is a complex quantity, reducing to $|{k}(\omega_i)| = \omega_i \,n/c_0$ with refractive index $n$ only for non-absorbing media.

\subsection{Experimental Results and Analysis}
\label{sec:spectrum}

\begin{figure}
	\centering
	\includegraphics[width=0.95\linewidth]{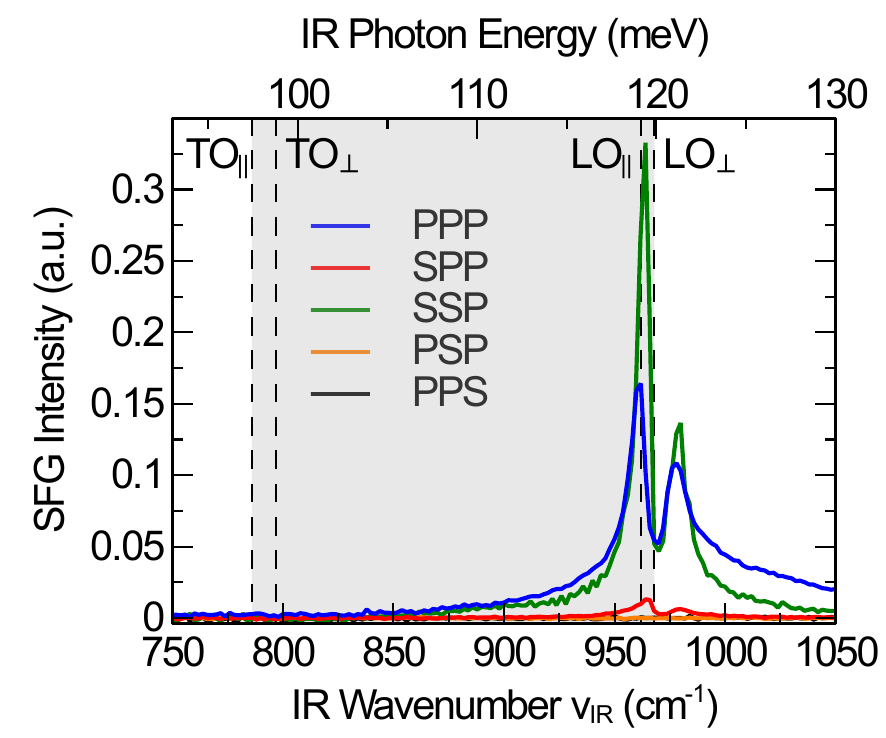}
	\caption{\label{fig:SiCexp}Experimental SFG spectra of 4H-SiC. The nonlinear response is depicted in dependence on the polarization of incoming and emitted waves within the Reststrahlen band (gray area), limited by the TO and LO phonon frequencies.}
\end{figure}

The static sum-frequency generation spectrum of the 4H-SiC crystal is obtained at zero time delay ($\tau = 0$) between IR excitation and VIS up-conversion pulse. 
Using the reflection geometry, SFG spectra under different  polarization combinations have been measured, summarized in Fig.~\ref{fig:SiCexp}. Here, the state of polarization of incoming and generated beams is denoted in the order $\hat{\b{e}}(\omega_3),\hat{\b{e}}(\omega_2),\hat{\b{e}}(\omega_1)$. Note that the spectra are plotted as function of the incident IR wavenumber $\nu_\t{IR}$.\\

Strong sum-frequency generation is observed in the cases of PPP and SSP polarization. There, a distinct double-peak structure of the SFG intensity arises around $\nu_\t{IR} \sim \SI{960}{cm^{-1}}$. In contrast, no significant SFG signal is measured at all under SPP, PSP and PPS conditions within the investigated spectral region of $\nu_\t{IR} = \SIrange{750}{1050}{cm^{-1}}$.\\

To rationalize these experimental findings, information about the number and symmetry properties of the vibrational modes are required. The characteristics of relevant optical phonons are summarized in Tab.~\ref{tab:sic}. Within the basal plane, the vibrational modes $\omega_{\t{TO/LO},\perp}$ at the $\Gamma$ point obey $E_1$ symmetry,
whereas the lattice oscillations $\omega_{\t{TO/LO},\parallel}$ polarized in the direction of the optic axis are $A_1$ symmetric~\cite{Mut99}. 
The $A_1$ and $E_1$ phonons are all Raman active~\cite{Hof94}. However, only the transverse vibrations can be excited by IR absorption~\cite{Pat68}.\\

At first sight, the measured spectral features might seem unreasonable, since at the position of the TO phonon, which is IR as well as Raman active and thus SFG-allowed~\cite{She16}, no peak is observed. On the other hand, the SFG intensity increase around the LO vibration can not be caused by a resonant enhancement of the second-order susceptibility due to the dipole-forbidden IR transition.\\

\begin{table}[t]
	\centering
	\caption{\label{tab:sic}Optical phonon frequencies of 4H-SiC at the Brillouin zone center in the Reststrahlen band. The values of phonon energy and damping are extracted from a fit of the SFG spectra using Eq.~(\ref{eq:SFGbulk}). Symmetry properties and IR / Raman activities are taken from Ref.~\cite{Hof94}. Planar modes ($\perp$) are IR active in the ordinary ray, axial modes ($\parallel$) are excited with the extraordinary ray.\smallskip}
	\begin{tabular}{cccccc}
		\hline \hline
		Mode & Symm. & $\nu$ (cm$^{-1}$) &  $\gamma$ (cm$^{-1}$) & IR-active & Raman-active\\
		\hline
		TO$_\parallel$ & $A_1$ & $787.2$ & $3.8$ & $\mathsf{x}$ & $\mathsf{x}$ \\
		LO$_\parallel$ & $A_1$ & $962.3$ &  &  & $\mathsf{x}$ \\
		TO$_\perp$ & $E_1$ & $797.6$ & $5.0$ & $\mathsf{x}$ & $\mathsf{x}$ \\
		LO$_\perp$ & $E_1$ & $968.5$ &  &  & $\mathsf{x}$ \\
		\hline \hline
	\end{tabular}
\end{table}

A fit of the spectra with the sum-frequency model developed in Eq.~(\ref{eq:SFGbulk}) enables to disentangle the different factors contributing to the SFG signal. Due to point group $6mm$ of the 4H-SiC crystal bulk~\cite{Hof94}, the second-order susceptibility tensor $\b{\chi}^{(2)}_\t{B}$ has the following four independent, non-vanishing components, given in Cartesian coordinates~\cite{Boy08}:
\begin{equation}
\label{eq:sic_chi2}
\chi^{(2)}_{zzz}, \chi^{(2)}_{zxx} = \chi^{(2)}_{zyy}, \chi^{(2)}_{xzx} = \chi^{(2)}_{yzy}, \chi^{(2)}_{xxz} = \chi^{(2)}_{yyz}.
\end{equation}

Within the investigated spectral range, the only resonant contribution to linear (Eq.~(\ref{eq:eps_sic})) and second-order (Eq.~(\ref{eq:FH})) optical susceptibilities arises due to a TO phonon mode, thus a single Lorentz term is sufficient for the description of the dispersion relation. The relevant vibrational modes $\omega_{\t{LO/TO},\,p}$ have to be considered in the model when the IR field polarization $\b{E}(\omega_1)$ is either parallel ($p = \parallel$) or perpendicular ($p = \perp$) to the $z$-axis.\\

Free fitting parameters are the phonon frequencies $\omega_{\t{LO/TO},\,p}$ and dampings $\gamma_{\t{TO},\,p}$, entering both into the dielectric function $\varepsilon_{p}(\omega)$ and the second-order susceptibility $\chi^{(2)}_{ijk}(\omega)$, as well as the non-resonant constants $\chi^{(2)}_{ijk\,\infty}$ and resonance amplitudes $A_{r,ijk}$ of the nonlinear susceptibility. The obtained phonon values are listed in Tab.~\ref{tab:sic}. Parameters of the dieletric function used in the IR region, Eq.~(\ref{eq:eps_sic}), are taken from Ref.~\cite{Mut99}, whereas Sellmeier equations, $\varepsilon_\perp(\omega) = n_\t{o}^2(\lambda)$ and $\varepsilon_\parallel(\omega) = n_\t{e}^2(\lambda)$, are employed in the VIS range~\cite{Wan13}.\\

\begin{figure}
	\centering
	\includegraphics[width=0.95\linewidth]{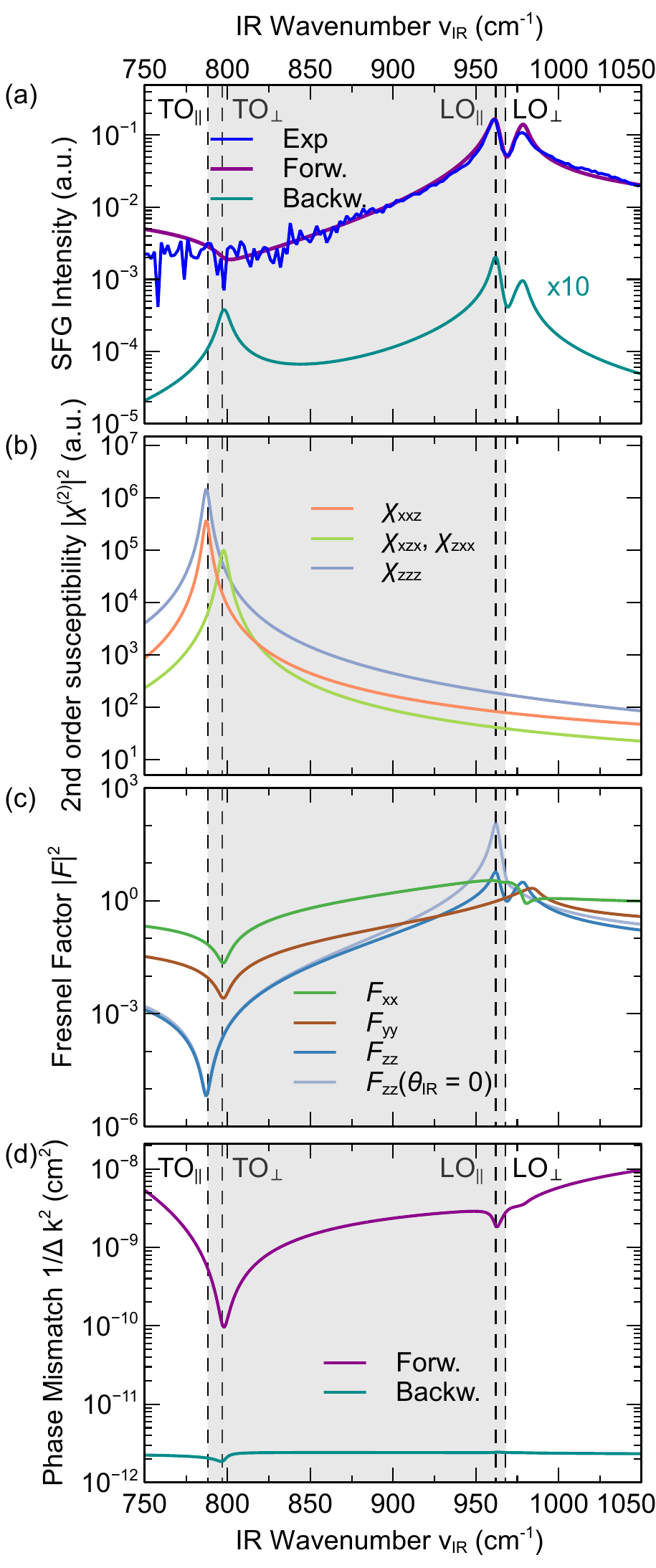}
	\caption{\label{fig:SiCtheo2}Sum-frequency generation in the Reststrahlen region of 4H-SiC. Shown are (a)~the total intensity as measured and fitted with the model of Eq.~(\ref{eq:SFGbulk}), (b)~the IR dispersion of the quadratic susceptibility tensor components, (c)~the Fresnel factors $|F_{ll}|^2$ governing the transmission of the IR field at the air/SiC interface and (d)~the phase mismatch between IR, VIS and SFG wave inside the nonlinear crystal. Depending on the propagation direction of the SFG radiation (forward, backward), the mismatch and hence the amplitude and spectral behavior of the emitted SFG light are affected. The broad range of the Reststrahlen band is indicated (gray area). Data are for PPP configuration.}
\end{figure}

The different contributions to the SFG spectrum in the case of PPP polarization, as extracted from the fit result, are displayed in Fig.~\ref{fig:SiCtheo2}. Experimental data and fitted curve show reasonable agreement, see Fig.~\ref{fig:SiCtheo2}(a). The nonlinear origin of the emitted radiation, captured by the dispersive $\b{\chi}^{(2)}$ tensor, is plotted with its absolute components $|\chi^{(2)}_{ijk}|^2$ in Fig.~\ref{fig:SiCtheo2}(b). Since the spectral variation of the up-converted radiation $\omega_3$ is rather small, only the IR frequency dependence of the second-order susceptibility $\chi^{(2)}(\omega_1)$ is taken into account.
In the case of PPP polarization, all independent tensor components contribute to the SFG signal. A resonant enhancement of the nonlinear susceptibility $\chi^{(2)}_{ijk}(\omega_1)$ of about two orders of magnitude is observed around the SFG-active TO phonon. Depending on the relevant tensor component, the peak occurs either at the axial ($\omega_{\t{TO},\parallel}$) or planar ($\omega_{\t{TO},\perp}$) TO mode.\\

To explain the difference of the peak positions in the measured SFG spectrum and the second-order susceptibility, linear optical effects have to be invoked~\cite{She08}. On the one hand, the Fresnel factors, Eq.~(\ref{eq:Fresneltensor}), depicted as $|F_{ll}(\omega_1,\theta_1)|^2$ in Fig.~\ref{fig:SiCtheo2}(c), modify the actual field strength within the material. For a $p$-polarized field $\b{E}(\omega)$, only the factors $F_{xx}$ and $F_{zz}$ are relevant. Reduced IR transmission at the crystal-air interface occurs for $\omega_1$ close to $\omega_\t{TO}$ due to the strong vibrational absorption of the crystal. Within the Reststrahlen band, the dielectric function is characterized by a negative real part. At the other boundary of the band, towards the LO vibration $\omega_\t{LO}$, the Fresnel factor is increased due to a near-zero permittivity. In particular, the out-of-plane component $F_{zz} E_z$ of the electric field rises by three orders of magnitude. Due to the optical anisotropy of the 4H-SiC crystal, both $\omega_{\t{LO}\,\parallel}$ and $\omega_{\t{LO}\,\perp}$ enter as singularities into $F_\t{zz}$, Eq.~(\ref{eq:Fresneltensor}), causing two spectrally separated peaks of the electric field enhancement. This effect is absent in the case of normal incidence ($\theta_1 = 0$), also shown in Fig.~\ref{fig:SiCtheo2}(c), where the crystal essentially shows an isotropic response.\\

On the other hand, the dielectric dispersion affects the wavevector mismatch $\Delta k(\omega_1)$ inside the nonlinear medium. Depending on the observation of the SFG radiation either transmitted or reflected from the crystal, the propagation of the sum-frequency wavevector $k_\t{z}(\omega_3)$ is either parallel to the incident beams $k_\t{z}(\omega_{1,2})$ (forward) or in opposite direction (backward). For both cases, the inverse square of the phase difference, $1/\Delta k^2$, as relevant quantity for the SFG intensity, cf. Eq.~(\ref{eq:SFGbulk}), is depicted in Fig.~\ref{fig:SiCtheo2}(d). Compared to the back traveling sum-frequency field, the wavevector mismatch $\Delta k$ of the SFG radiation in forward propagation direction through the transparent crystal is rather small.
Thus, the intensity of the forward contribution dominates the measured SFG signal, Fig.~\ref{fig:SiCtheo2}(a), due to reflection at the optically flat backside of the crystal (reflectance at SiC/air interface $|r_\t{p}|^2 = 0.16$) towards the detector. The IR dispersion is less pronounced in the phase mismatch of the backward direction, Fig.~\ref{fig:SiCtheo2}(a), because of $|k_\t{z}(\omega_1)|\ll |k_\t{z}(\omega_2)|$ and $\Delta k_\t{z} \approx 2 k_\t{z}(\omega_2)$. These wave vector mismatch values result in effective probe depths  in the order of microns and $\approx$100~nm~\cite{Hor19} for forward and backward SFG emission, respectively.\\

Consequently, the linear optical effects of Fresnel transmission and wavevector mismatch strongly modify the SFG spectrum. The contribution of the resonant enhancement of $\chi^{(2)}(\omega_1)$ at the TO phonon frequency $\omega_\t{TO}$ is suppressed by the reduced electric field strength $\bu{F}(\omega_1) \b{E}(\omega_1)$ and increased phase-mismatch $\Delta k(\omega_1)$. Instead, a (double) peak of the SFG intensity is induced at the LO mode $\omega_\t{LO}$, mainly caused by the enhanced Fresnel transmission of the out-of-plane component of the electric field $F_\t{zz} E_\t{z}$. A similar behavior is observed for the SFG spectra at other polarization conditions with non-vanishing $\chi^{(2)}$ components (Fig.~\ref{fig:SiCexp}).\\
\\

\begin{figure*}
	\centering
	\includegraphics[width=0.95\linewidth]{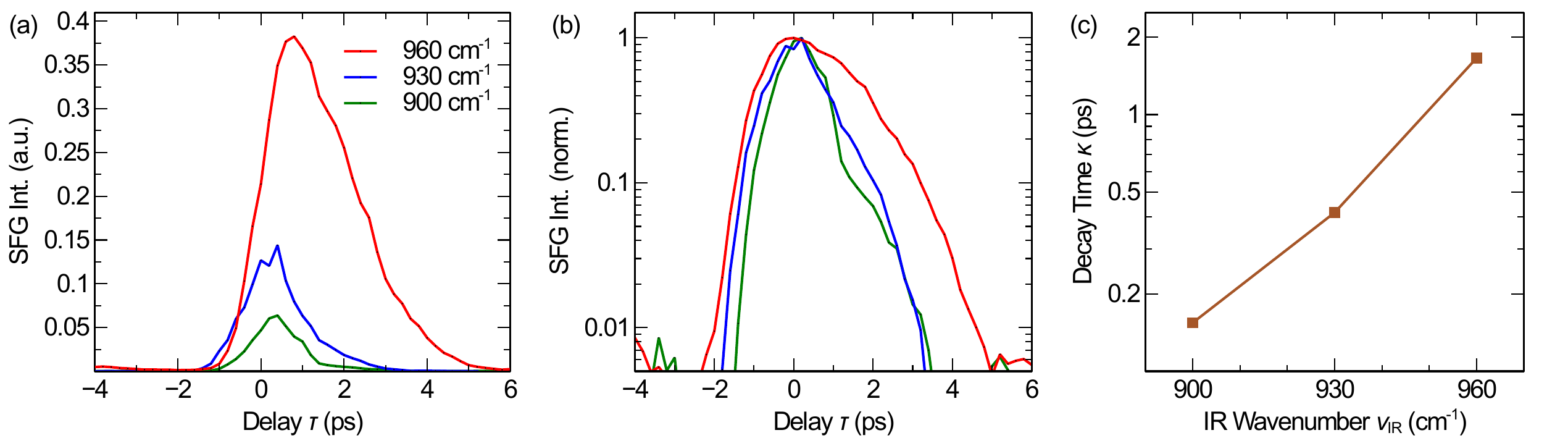}
	\caption{\label{fig:trSFG}Free-induction decay of SFG intensity. (a) SFG time-domain traces for different IR excitation wavenumbers $\nu_\t{IR}$ close to the LO phonon mode. Time axis is given as the delay between IR excitation and VIS up-conversion pulse. (b) Normalized decay transients of plot (a) on a semi-logarithmic scale. (c) Decay constants $\kappa$ at different IR excitation wavenumbers. Each constant is extracted from a convolution fit of Eq.~(\ref{eq:conv}) and a reference excitation pulse shape to the SFG 2D graphs (see Fig.~\ref{fig:2Dtraces}). The dephasing time is related to the decay constant via $T_2 = 2 \kappa$.}
\end{figure*}

From the experiment, only the ratio of the susceptibility tensor components $\chi^{(2)}_{ijk}$ can be extracted. The element $|\chi^{(2)}_\t{zzz}|$ is stronger than the other non-vanishing components, cf. Fig.~\ref{fig:SiCtheo2}(b). In comparison to reported experimental values for the quadratic nonlinear optical coefficients of 4H-SiC in the IR region, $\chi^{(2)}_\t{zzz}/\chi^{(2)}_\t{zxx} = -4.5$~\cite{Nie99} and $|\chi^{(2)}_\t{zxx}| \approx |\chi^{(2)}_\t{xxz}| \approx \SI{6.5}{pm/V}$~\cite{Sat09}, the SFG measurement is in reasonable agreement. Calculated second-order susceptibilities, based on density functional theory, support the experimental results~\cite{Che94,Hue11}. Moreover, it has been found that the optical nonlinearity is increased for SiC polytypes of reduced hexagonality (e.g. 6H, 3C), which possess also a smaller band gap~\cite{Nie99}.

\section{Time-Domain Sum-Frequency Dynamics}

\subsection{Model}
\label{sec:tdmodel}

In time-domain SFG measurements, the control of the temporal delay $\tau$ of the IR pulse excitation $\b{E}_\t{IR}$ versus the VIS up-conversion field $\b{E}_\t{VIS}$ provides insights into the dephasing dynamics of the induced coherent IR polarization $\b{P}^{(1)}(t)$~\cite{Ueb97}. Thus, by recording the SFG intensity as a function of the delay, $I_\t{SFG}(\tau)$, the free-induction decay (FID) of the linear polarization $\b{P}^{(1)}(t)$ can be captured~\cite{Bon00}.\\

An analytical model of the time-domain SFG process allows to calculate the signal intensity $I_\t{SFG}(\tau)$. Starting from the IR pulse $E_\t{IR}(t)$, the interaction with the investigated system, characterized by a response function $R(t)$, causes a first-order polarization $\b{P}^{(1)}(t)$~\cite{Bor05,She16}:
\begin{equation}
	P^{(1)}(t) = \int \diff t' E_{\t{IR}} (t-t') R(t').
\end{equation}

In case of resonant excitation of the system, a long-lasting coherent polarization is induced, dephasing over time with a time constant $T_2$. The delayed up-conversion pulse $E_\t{VIS}(t - \tau)$ creates the nonlinear polarization $\b{P}^{(2)}(t)$,
\begin{equation}
	\label{eq:P2}
	P^{(2)}(t,\tau) \approx E_\t{VIS} (t-\tau) P^{(1)}(t).
\end{equation}

Due to the non-resonant field $E_\t{VIS}$, an instantaneous interaction with the linear polarization $P^{(1)}(t)$ can be assumed, thus not requiring a convolution integral. Finally, the SFG intensity emitted by the second-order polarization,
\vspace{-3mm}
\begin{equation}
	\label{eq:Isfg}
	I_\t{SFG}(\tau) \propto \int \diff t |P^{(2)}(t,\tau)|^2,
\end{equation}
is obtained as a function of the time delay $\tau$.

\subsection{Free-Induction Decay of IR Polarization}

\begin{figure*}
	\centering
	\includegraphics[width=0.95\linewidth]{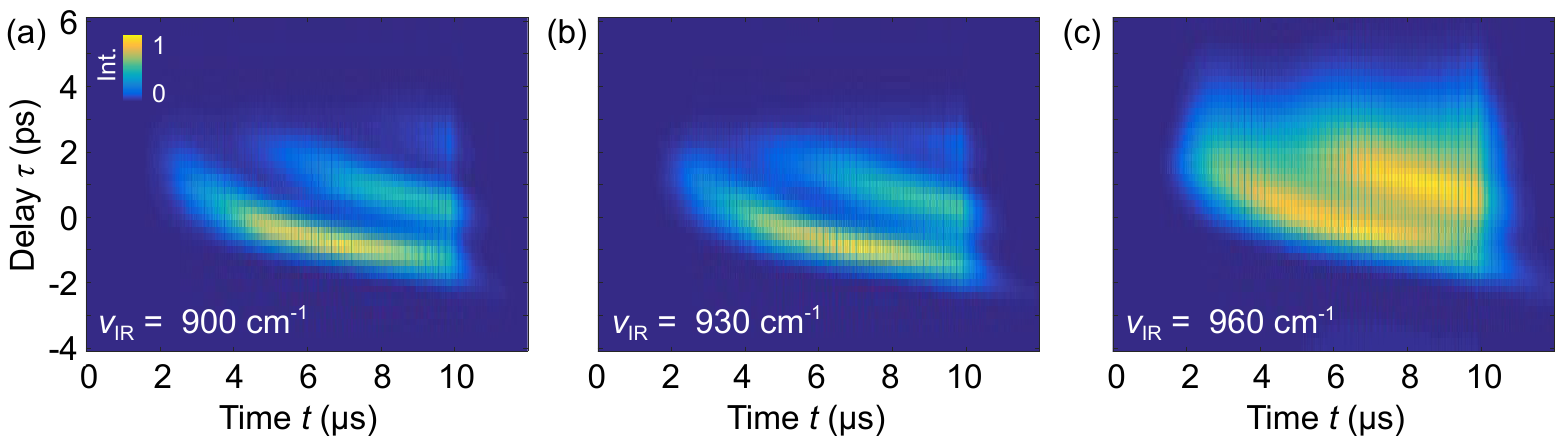}
	\caption{\label{fig:2Dtraces}Two-dimensional graphs of the time-domain SFG response. The delay-resolved free-induction decay $I_\t{SFG}(\tau)$ is shown during the full FEL macro-pulse real-time $t$. Different IR excitation wavenumbers of (a)~$\nu_\t{IR} = \SI{900}{cm^{-1}}$, (b)~$\SI{930}{cm^{-1}}$ and (c)~$\SI{960}{cm^{-1}}$ are used. The multi-peak structure along the delay axis $\tau$ is caused by the sub-pulse emission within a single micro-pulse of the IR FEL. The subsequent decay dynamics of the SFG intensity shows the dephasing of the induced polarization as the response of the material. In all cases, short IR micro-pulses generated at FEL cavity detuning $\Delta L = 1 \lambda$ are used.}
\end{figure*}

The dephasing of the induced dielectric polarization $P^{(1)}$ in 4H-SiC has been studied close to the LO phonon mode. Although the vibration at $\nu_{\t{LO}}$ is not an SFG-active $\chi^{(2)}$ resonance, the phonon-related modulation of the linear optical response $R=F(\nu)$ provides for a strong SFG signal, as discussed previously in Sec.~\ref{sec:spectrum}. All time-domain measurements have been performed in PPP polarization configuration. For the IR excitation, a small FEL cavity detuning length $\Delta L = 1 \lambda$ was employed in order to generate FEL pulses with a duration of $\lesssim 1$~ps (cf. Fig.~\ref{fig:2Dtraces}), in contrast to few-ps pulses used in the frequency-domain SFG measurements. The SFG transients $I_\t{SFG}(\tau)$ obtained at different excitation frequencies $\nu_\t{IR}$ around $\nu_{\t{LO}} \sim \SI{960}{cm^{-1}}$ are plotted in Figs.~\ref{fig:trSFG}(a),(b), after BOC correction of the delay values $\tau$, see Appendix for details.\\

Clear differences in the behavior of the IR polarization $P^{(1)}(t)$ can be derived from the SFG graphs $I_\t{SFG}(\tau)$, Fig.~\ref{fig:trSFG}(a): The spectral dependence of the peak amplitude arises from the Fresnel factor $F(\nu)$, i.~e. linear optical effects. With regard to the dynamics, a slower free-induction decay $P^{(1)}(t)$ is observed for resonant excitation at $\nu_\t{IR}= \SI{960}{cm^{-1}}$, as compared to the off-resonant traces at $\nu_\t{IR}=\SI{900}{cm^{-1}}$ and $\nu_\t{IR}=\SI{930}{cm^{-1}}$. This becomes obvious from the normalized semilogarithmic plots in Fig.~\ref{fig:trSFG}(b). Further, a small temporal shift of the peak position by about $\SI{400}{fs}$ towards positive delay times is detected in the resonant case, cf. Fig.~\ref{fig:trSFG}(a).\\

In order to compare the dephasing dynamics at the different excitation frequencies $\nu_\t{IR}$, decay constants are extracted from the experimental data.
To this end, a fit of the measured transients is performed with the convolution of a reference FEL pulse shape and a step-like exponential function of the form
\vspace{-0.5mm}
\begin{equation}
	\label{eq:conv} 
	f(\tau) = H(\tau) \, \eu^{-\tau/\kappa},
\end{equation}
where $H(\tau)$ is the Heaviside distribution. The resulting SFG intensity decay times $\kappa$ are displayed in Fig.~\ref{fig:trSFG}(c).\\

The IR excitation in the vicinity of the LO phonon mode reveals a slower decay constant of $\kappa = 1.66 \pm \SI{0.01}{ps}$, compared to the significantly faster decay for the spectrally offset case at $\nu_\t{IR}= \SI{900}{cm^{-1}}$ with $\kappa = 0.15 \pm \SI{0.01}{ps}$. Since the SFG intensity is related to the linear polarization via $I \propto |P^{(2)}|^2$ and $P^{(2)} \propto P^{(1)}$, the dephasing time $T_2$ of the IR polarization amounts to $T_2 = 2 \kappa$~\cite{Hes02}. Thus, the measured free-induction decay time in bulk 4H-SiC varies between $T_2 = 3.32 \pm \SI{0.02}{ps}$ (at the LO phonon resonance) and $T_2 = 0.30 \pm \SI{0.02}{ps}$ (off-resonant).
Although the LO vibration itself does not absorb the IR light, the presence of the vibrational mode affects the dephasing of the IR polarization through the dispersive linear-optical behavior. For comparison, dephasing times of the same order of magnitude have been found in solids for the LO phonon ($T_2 = 4.2 \pm \SI{0.4}{ps}$~\cite{Val94}) and TO phonon ($T_2 = 2.9 \pm \SI{0.3}{ps}$~\cite{Gan97}) in a semiconducting polar GaAs crystal or the TO mode ($T_2 = 5.8 \pm \SI{0.6}{ps}$~\cite{Lau71}) in diamond at room temperature.\\
 
The decay dynamics of the IR polarization can also been visualized using the full FEL pulse profile, see Fig.~\ref{fig:2Dtraces}. As reported recently, Ref.~\cite{Kie18}, the infrared emission of an FEL oscillator can exhibit a sub-pulse structure under certain conditions. There, multiple intensity peaks are present during the ps-short FEL micro-pulse, which give rise to a regular oscillation pattern over the course of the $\mu$s-long IR macro-pulse (see Fig.~5 in Ref.~\cite{Kie18}). The incorporation of the sub-pulse dynamics of the FEL into the measured dephasing response of the solid material is depicted in Figs.~\ref{fig:2Dtraces}(a)-(c). Here, the SFG intensity $I_\t{SFG}(\tau)$ is plotted as function of IR-VIS delay (corrected for timing jitter, see Appendix) and real time $t$ within an FEL macro-pulse for different IR excitation wavenumbers $\nu_\t{IR}$.\\

In Fig.~\ref{fig:2Dtraces}(a), the temporal structure of the SFG signal closely follows the original FEL sub-pulse pattern, due to the fast dephasing in the non-resonant case at $\nu_\t{IR}=\SI{900}{cm^{-1}}$. At the beginning of the macro-pulse, at about $t=\SI{2}{\micro s}$, the first sub-pulse emerges. In the course of time $t$, the peak position is shifted towards a negative delay $\tau$. Meanwhile, the second sub-pulse arises at $t \sim \SI{4}{\micro s}$, undergoing the same delay shift. The commencing of a third sub-pulse is observed near the end of the macro-pulse at $t=\SI{10}{\micro s}$. These observations of the sub-pulse behavior could be well reproduced by FEL simulations based on classical electrodynamics~\cite{Kie18,Col90}.\\

The variation of the dephasing dynamics with the excitation frequency is apparent in Figs.~\ref{fig:2Dtraces}(b) and (c), when looking at the SFG intensity contrast in the temporal range between the sub-pulse peaks. In the resonant case at $\nu_\t{IR}=\SI{960}{cm^{-1}}$, Fig.~\ref{fig:2Dtraces}(c), a clear 'smear-out' of the intensity is oberved, caused by the decay time $\kappa$ being signficantly longer than the sub-ps FEL micropulses, compared to the near-instantaneous response leading to much `crisper' features in Figs.~\ref{fig:2Dtraces}(a) and (b). The one-dimensional dephasing traces shown in Fig.~\ref{fig:trSFG} have been extracted from the plots of Fig.~\ref{fig:2Dtraces} at time $t=\SI{3.5}{\micro s}$, to facilitate the fitting of the material response induced by a single FEL sub-pulse.

\subsection{Transient SFG Spectra}

\begin{figure}
	\centering
	\includegraphics[width=0.8\linewidth]{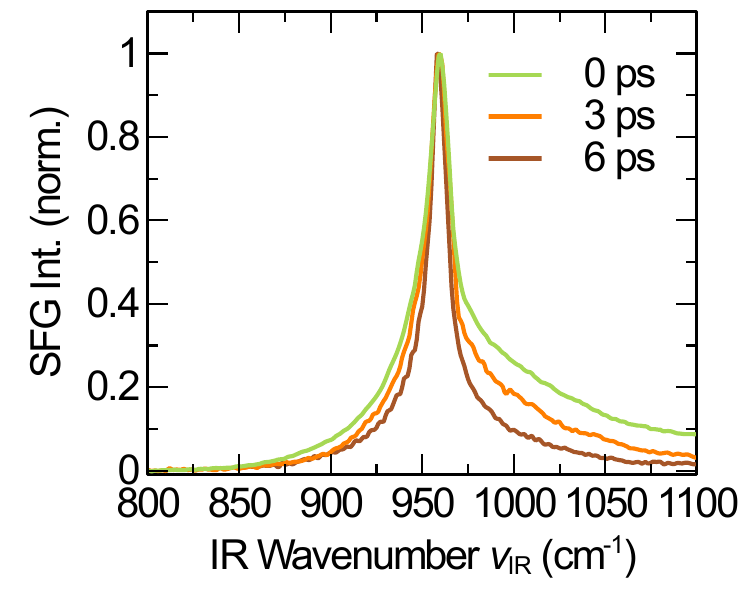}
	\caption{\label{fig:trspectra}Transient SFG spectra near LO phonon resonance. The normalized SFG intensity is shown at distinct time delays~$\tau$ between the excitation with narrowband IR FEL radiation ($\Delta L = 4 \lambda$) and the subsequent up-conversion pulse.}
\end{figure}

A further characterization of the polarization dephasing has been carried out in the frequency domain by acquiring SFG spectra $I_\t{SFG}(\nu_\t{IR})$ at different delay times $\tau$, see Fig.~\ref{fig:trspectra}. Here, the spectral response is acquired by scanning the wavenumber $\nu_\t{IR}$ of the narrowband FEL radiation (cavity detuning $\Delta L = 4 \lambda$) while keeping the delay path relative to the up-conversion pulse constant.\\

For all measured delays in Fig.~\ref{fig:trspectra}, a single resonance peak around $\nu_\t{IR}=\SI{960}{cm^{-1}}$ is present. This is in contrast to the static SFG spectrum measured in reflection geometry, Fig.~\ref{fig:SiCtheo2}(a), where two peaks have been observed. The reason is the smaller angle of incidence of the IR field ($\sim \SI{25}{\degree}$) used in the transient SFG spectra measurements in transmission geometry, modifying the angle-dependent Fresnel factor $F(\nu_\t{IR},\theta_\t{IR})$,
as illustrated for normal incidence in Fig.~\ref{fig:SiCtheo2}(c).\\

The linewidth extracted from the spectrum at $\tau = 0$ amounts to $21.8\pm\SI{0.7}{cm^{-1}}$ (FWHM). Compared to the time-domain data ($T_2 = \SI{3.32}{ps}$), a much smaller vibrational linewidth $2 \Gamma = 1/ \pi c_0 T_2 = \SI{3.2}{cm^{-1}}$ would be expected for a homogeneously broadened Lorentzian lineshape~\cite{Lau78}. 
In fact, the observed resonance has an asymmetric, non-Lorentzian profile, cf. Fig.~\ref{fig:trspectra}. As discussed previously, the peak in the SFG spectra around $\nu_\t{IR} \sim \SI{960}{cm^{-1}}$ does not originate from a resonant enhancement of the second-order susceptibility $\chi^{(2)}(\nu_{\t{IR}})$, which would result in a Lorentzian, but is due to a frequency-dependent modulation of the nonlinear response by the Fresnel factor $F(\nu_{\t{IR}})$. In consequence, the free-induction decay traces in the time-domain deviate from an ideal exponential function, cf. Fig.~\ref{fig:trSFG}. Nevertheless, the fitting result provides a comparative figure.\\

The development of the normalized spectral lineshape $I_{\t{SFG}}(\nu_\t{IR})$ as a function of the temporal delay $\tau$ in Fig.~\ref{fig:trspectra} exhibits two main characteristics: First, the wavenumber of the peak position at $\hat{\nu} \approx \SI{960}{cm^{-1}}$, related to the LO phonon $\nu_{\t{LO}}$, remains constant over time. Second, a narrowing of the spectral linewidth $\Delta \nu$ is observed at a larger time delay $\tau$ between the IR excitation and VIS up-conversion pulse. This is accompanied by a transition of the lineshape from the asymmetric to a more symmetric form. The extracted peak wavenumbers $\hat{\nu}$ and bandwidths $\Delta \nu$ (FWHM) of the resonance are summarized in Tab.~\ref{tab:linefwhm}. These spectral observations are in line with the frequency-dependent dephasing measurements in the time-domain, cf. Fig.~\ref{fig:trSFG}. The faster off-resonant decay of the induced IR polarization entails a spectral narrowing of the linewidth towards longer delay values. We note that spectral narrowing in different time-domain spectroscopy approaches has been predicted to even enable surpassing the resolution limit given by the dephasing time of the free induction decay~\cite{Zin83}.  

\begin{table}[t]
	\centering
	\caption{\label{tab:linefwhm}Development of peak frequency $\hat{\nu}$ and bandwidth $\Delta \nu$ of the SFG resonance as function of the IR-VIS pulse delay $\tau$. The values are extracted from the transient spectra shown in Fig.~\ref{fig:trspectra}.\smallskip}
	\begin{tabular}{cccc}
		\hline \hline
		Delay $\tau$ & $\SI{0}{ps}$ & $\SI{3}{ps}$ & $\SI{6}{ps}$ \\
		\hline
		Peak $\hat{\nu}$ (cm$^{-1}$) & $959.3 \pm 0.2$ & $959.3 \pm 0.1$ & $958.9 \pm 0.1$ \\ 
		FWHM $\Delta \nu$ (cm$^{-1}$) & $21.8 \pm 0.7$ & $18.1 \pm 0.5$ & $13.1 \pm 0.3$\\ 
		\hline \hline
	\end{tabular}
\end{table}

\subsection{Simulation}

To support the experimental results, simulations of the transient SFG response have been performed. Following the theoretical model in Sec.~\ref{sec:tdmodel}, the temporal dephasing of the material's first-order polarization $\b{P}^{(1)}(t)$, induced by the IR excitation, is studied. The response function $R(\nu)$ of the SiC crystal in the frequency domain is characterized by the Fresnel factor $\b{F}(\theta, \nu)$, considering the resonance near the LO phonon mode. Taking into account a Gaussian pulse shape, the IR pulse can be described by a variable center wavenumber $\nu_0$ and a fixed spectral width $\sigma_\nu = 0.441 /2 \sqrt{2 \ln 2} \tau_\t{IR}$ related to the pulse duration $\tau_\t{IR}$,
\vspace*{-2mm}
\begin{equation}
	\label{eq:Eir}
	\b{E}_\t{IR}(\nu) = \hat{\b{E}} \eu^{-\frac{(\nu - \nu_0)^2}{2\sigma^2_\nu}}.
\end{equation}
The resulting linear polarization, given by
\begin{equation}
	P^{(1)}(\nu) \propto |\b{F}(\nu) \b{E}_\t{IR}(\nu)|,
\end{equation}
is subsequently Fourier-transformed to obtain the free-induction decay in the time-domain, 
\begin{equation}
	\label{eq:Pirt}
	P^{(1)}(t) = \mathcal{F}(P^{(1)}(\nu)).
\end{equation}
Insertion of the IR polarization $P^{(1)}(t)$ into Eqs.~(\ref{eq:P2}) and (\ref{eq:Isfg}) yields the time-evolution of the SFG intensity, $I_\t{SFG}(\tau)$. The spectral dependence $I_\t{SFG}(\tau,\nu)$ is calculated by tuning of the IR wavenumber $\nu_0$ in Eq.~(\ref{eq:Eir}).\\

\begin{figure*}
	\centering
	\includegraphics[width=0.82\linewidth]{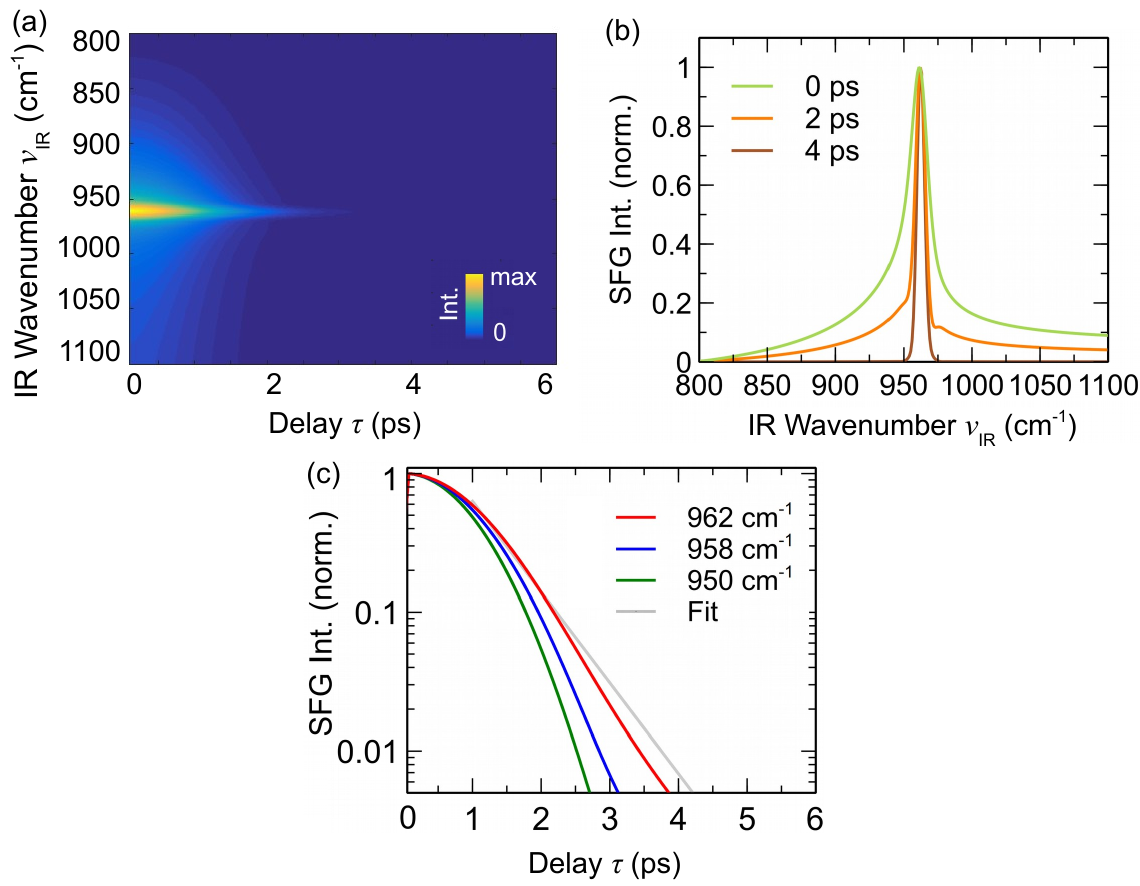}
	\caption{\label{fig:resmodel}Simulation of the time-domain SFG response of a single resonance. (a)~Two-dimensional plot of the SFG intensity $I_\t{SFG}(\tau,\nu_\t{IR})$, showing a decay of the intensity due to polarization dephasing. (b)~Transient normalized SFG spectra $I_\t{SFG}(\nu_\t{IR})$ extracted at certain time delays $\tau$. The single resonance is located around $\nu_0 = \SI{962}{cm^{-1}}$. (c)~Free-induction decay of the SFG intensity $I_\t{SFG}(\tau)$, depicted for different IR excitation wavenumbers $\nu_\t{IR}$. An exponential fit to the trace at $\nu_0 = \SI{962}{cm^{-1}}$ is also shown (gray line).}
\end{figure*}

The simulation results of the time- and frequency-dependent SFG intensity are depicted in Fig.~\ref{fig:resmodel}. The chosen calculation conditions are a close to normal incidence of the IR excitation with $\theta_{\t{IR}}=\SI{25}{\degree}$, a pulse duration of $\tau_{\t{IR}} = \SI{2}{ps}$ and a PPP polarization configuration, corresponding to the measurement situation. In Fig.~\ref{fig:resmodel}(a), the two-dimensional plot shows a single resonance, located around $\nu_{\t{IR}}=\SI{962}{cm^{-1}}$, decaying in intensity on a ps-time scale. The normalized SFG spectra $I_\t{SFG}(\nu_\t{IR})$ at certain time delays~$\tau$ are displayed in Fig.~\ref{fig:resmodel}(b). On the one hand, an asymmetric lineshape is present at zero delay. Towards larger delay times, the resonance profile turns symmetric. On the other hand, a narrowing of the spectral linewidth is observed at increased pulse delays, e.g. to $\SI{32}{\percent}$ of the original FWHM after $\tau = \SI{4}{ps}$. These theoretical findings are in qualitative agreement with the measured behavior, cf.~Fig.~\ref{fig:trspectra}.\\ 

In Fig.~\ref{fig:resmodel}(c), the simulated temporal development of the SFG intensity $I_\t{SFG}(\tau)$ is depicted for different excitation wavenumbers $\nu_{\t{IR}}$. There, the intensity reflects the free-induction decay of the induced linear polarization $P^{(1)}(t)$. Obviously, an off-resonant IR excitation yields a faster dephasing than in the resonant case. For excitation at the LO phonon wavenumber, $\nu_{\t{IR}}=\SI{962}{cm^{-1}}$, a decay time of $\kappa=\SI{0.7}{ps}$ might be extracted from an exponential fit. However, the calculated SFG trace deviates from the exponential form, see Fig.~\ref{fig:resmodel}(c). This is in accordance with the non-Lorentzian lineshape in the calculations, Fig.~\ref{fig:resmodel}(b). Also, the experimental SFG time-domain traces in Fig.~\ref{fig:trSFG}(b) exhibit a slight deviation from the exponential decay. Consequently, the applied model simulations provide a valuable description of the experimental observations.

\section{Conclusion}

In this work, combined time- and frequency-domain IR-VIS sum-frequency generation spectroscopy has been performed on vibrational resonances in a semi-insulating, non-centrosymmetric crystal of silicon carbide. Whereas only phonon modes of both IR- and Raman-activity are allowed for a resonant enhancement of the second-order susceptibility $\chi^{(2)}$, strongly dispersive linear-optical effects such as the Fresnel transmission coefficients are found to significantly modify the SFG spectra. In the time-domain, a free-induction decay of the SFG intensity on a ps-timescale is detected, caused by the dephasing of the induced coherent IR polarization. Tuning to the phonon resonance, a slowdown of the dephasing time $T_2$ has been observed. As a consequence, using a temporal delay between the excitation and up-conversion pulse, spectral narrowing of the resonance could be demonstrated. The experimental results are supported by model simulations of the transient SFG response.\\

Time-domain SFG spectroscopy studies have shown a higher sensitivity to the resonant part in the SFG process if IR and VIS pulses are temporally separated, avoiding the non-resonant, instantaneous contribution of the second-order susceptibility~\cite{Rok03,Lag07,Laa11}. Thus, the lineshape in the frequency domain is largely determined by the resonance dephasing, providing a reasonable way to enhance the spectral resolution~\cite{Zin83}. 

\begin{acknowledgments}
	
The authors are grateful to S. Gewinner and W. Sch\"{o}ll-\\
kopf for reliable operation of the free-electron laser.
 
\end{acknowledgments}

\appendix*
\section{Timing Correction}
\label{sec:appendix}

The implementation of the balanced optical cross-correlator (BOC) as well as the characterization of the timing jitter and drift at the employed FEL has been reported previously~\cite{Kie18}. Here, the BOC is used as timing monitor in parallel with the time-resolved measurement to compensate for drifts by subsequent post-processing of the acquired data. For that, the timing value $\delta \tau$ received from the calibrated BOC is added to the delay setting $\tau$ to obtain the actual timing $\tau + \delta \tau$ for each FEL macro-pulse. Subsequently, the SFG intensity at the original delay point, $I(\tau)$, is retrieved by linear interpolation and averaging over multiple ($\sim 20$) macro-pulses per delay point.\\

\begin{figure}
	\centering
	\includegraphics[width=0.98\linewidth]{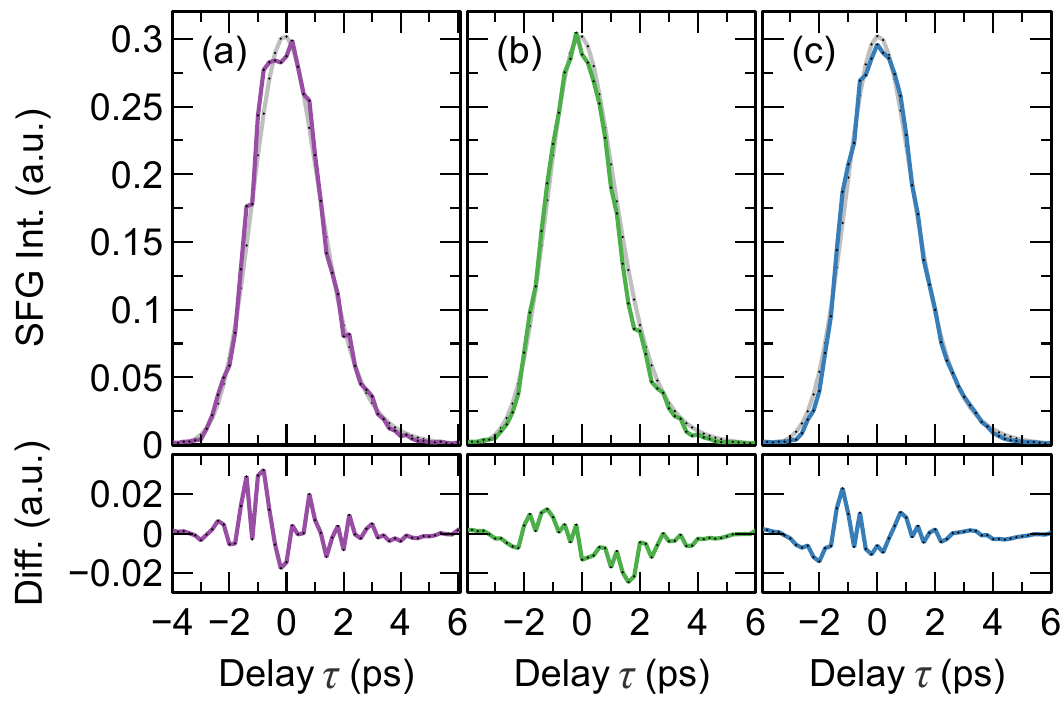}
	\caption{\label{fig:corrMethod}Post-correction of FEL pulse timing. The time-domain SFG intensity trace, $I_\t{SFG}(\tau)$, is shown in the upper panel (a) as measured and after application of the delay correction using the (b) kinetic electron energy monitoring (BPM) or the (c) balanced optical cross-correlation (BOC), respectively. Displayed traces are obtained after timing correction on a macro-pulse level and subsequent averaging. The quality of the results is assessed by comparing with the best-fit to the BOC-corrected SFG transient in (c) using Eq.~(\ref{eq:exgauss}), depicted as gray line in (a) to (c). Lower panel shows the calculated intensity differences between data and fit.}
\end{figure}

The results of the timing correction applied to the time-domain SFG measurement are shown in Fig.~\ref{fig:corrMethod}. While the raw data are displayed in Fig.~\ref{fig:corrMethod}(a), the BOC-adjusted SFG trace is plotted in Fig.~\ref{fig:corrMethod}(c). To quantify the improvement in the data quality after correction, a model function is fitted to each curve and the root-mean-square deviation (RMSD) is calculated. The model function, a convolution integral of a Gaussian pulse shape with an exponentially decaying material response, is 

\begin{multline}
	\label{eq:exgauss}
	I(\tau) = I_0 \exp \left( \frac{\sigma^2}{2\kappa^2} - \frac{\tau - \tau_0}{\kappa} \right) \\
	\left( 1 - \t{erf} \left( \frac{ \sigma^2 - \kappa(\tau - \tau_0) }{ \sqrt{2}\sigma\kappa } \right) \right),
\end{multline}
with the pulse width $\sigma$, decay constant $\kappa$ and delay zero at $\tau_0$. The reduced RMSD of the BOC-corrected data, $\SI{6.5e-3}{}$, compared to the uncorrected measurement, $\SI{9.6e-3}{}$, reveals the improvement in the data quality.\\

In addition, the electron beam-position monitoring (BPM) of the accelerator-driven FEL has been employed for timing correction, see Fig.~\ref{fig:corrMethod}(b). Due to the  correlation between kinetic electron energy and pulse arrival time (cf.~Ref.~\cite{Kie18}) with a correlation coefficient of $\rho=0.65$, a RMSD of the BPM-corrected curve of $\SI{8.2e-3}{}$ could be achieved. Thus, a post-processing of the FEL pulse timing merely based on the monitoring of the electron beam characteristics might provide an alternative method when the optical cross-correlation is difficult to obtain (e.g., due to the used FEL wavelength)~\cite{Cav05}. However, the correction accuracy is usually lowered compared to the all-optical timing tool.

\end{document}